\documentclass{iopbk2e}
\usepackage{graphics}
\usepackage{subfigure}
\usepackage{amssymb}
\usepackage{myapjbib}

\begin{document}
\title{Stars and singularities:\\
 Stellar phenomena near a massive black hole}
\author{Tal Alexander}
\date{
To appear in\\
{\Large \bf \rule{0em}{1.5em} The Galactic Black Hole}\\
\rule{0em}{1.5em}
H. Falcke \& F. W. Hehl, eds.\\
\rule{0em}{1.5em}
Institute of Physics 2002}
\maketitle
\pagenumbering{roman}
\setcounter{page}{1}
\tableofcontents
\newpage

\pagenumbering{arabic}
\setcounter{page}{1}

\Chapter[Stars and singularities]
{Stars and singularities: Stellar
phenomena near a massive black hole}{
{Tal Alexander}\\
{The Weizmann Institute of Science}\\}
\label{chap:Stars}

\section{Introduction}

\label{sec:Intro}

Isolated black holes are simple objects, characterized by three numbers
only: mass, angular momentum and charge. The complexity arises from
the interaction with their surroundings, which results in a wealth
of physical phenomena. This chapter will focus on the interaction
of the central \( \sim \! 3\times 10^{6}\, M_{\odot } \) massive
black hole (MBH) in the Galactic Center (Genzel et al. \cite{Gen00})
with the stars very close to it. It will discuss processes for which
there is already some observational evidence, as well as processes
that are suggested by theory and may yet be discovered by future observations
of the MBH in the Galactic Center or of central MBHs in other galaxies.

The MBH environment is unique because in addition to the gravitational
singularity, there are three other {}``effective singularities''
that are associated with the MBH. (1) A stellar density singularity.
This is predicted to occur in most scenarios for the evolution of
a stellar system around a MBH (e.g. Bahcall \& Wolf \cite{Bah76},
\cite{Bah77}; Young \cite{You80}). A density distribution that formally
diverges at the origin is called a cusp. In practice, infinite density
is not reached. Stars cannot exist closer than the event horizon,
and in fact they are destroyed well before that point either by collisions
or by the MBH tidal field. (2) A velocity singularity. Close to the
MBH the velocity field is Keplerian and so formally diverges as \( r^{-1/2} \).
The velocity cusp is also limited in practice by the absence of stars
arbitrarily close to the MBH. (3) An optical singularity. Any mass
bends light and amplifies the flux of background sources. Behind the
MBH (or any other sufficiently compact mass) there is a small region
(a caustic) where the amplification formally diverges to infinity.
This divergence is truncated by the finite size of the source.

The discussion will focus on the consequences of these singularities
on stars near the MBH, where the term {}``near'' is defined here
to mean the region where stars can exist (i.e. beyond the tidal radius)
but where the potential is completely dominated by the MBH. For the
Galactic Center, the event horizon is much smaller than the tidal
radius (for a solar type star) and so General Relativistic effects
can be neglected to first approximation.

There are several reasons to study stars near the MBH. First, unlike
gas, whose dynamics can be influenced by non-gravitational forces
such as thermal pressure, radiation pressure and magnetic fields,
stars are clean gravity probes. The properties of stars are well known
from other environments, and their observed luminosity and spectrum
can be translated into mass and maximal age. Both these quantities
are very important for understanding the dynamics of the system. In
particular, processes that operate on time-scales much longer than
the maximal stellar age cannot be relevant for the star. Second, stars
very near the MBH are connected to the growth of the MBH through tidal
disruption, mass loss from stellar winds and from stellar collisions.
Third, the region near the MBH can provide a unique laboratory for
studying stellar phenomena under extreme conditions: high density,
velocity and strong tidal fields.

Presently, infrared spectroscopy is possible for the brighter, well
separated stars in the field. Spectroscopy indicates that the stellar
population is a mix of old (red) stars and young (blue) stars (see
review by Genzel, Hollenbach \& Townes \cite{Gen94}). The old red
giants seen near the MBH in the Galactic Center are in the mass range
\( \sim 1 \)--\( 8M_{\odot } \) and are older than 1 Gyr. The faintest
observable young blue giants may be main sequence B1 or O9 stars with
masses of \( \sim 20M_{\odot } \) and main-sequence lifetimes \( <5 \)
Myr. The brightest young stars, the {}``He stars'', are Wolf Rayet-like
stars with masses of \( >20M_{\odot } \) and lifetimes of \( <10 \)
Myr. The blue stars are too young to have relaxed dynamically, and
their orbits (position, velocity) still reflect the initial conditions
of their formation (e.g. the young blue emission line giants are observed
to counter rotate relative to the galactic rotation).

All the stars in the inner 0.02 pc around SgrA\( ^{\star } \) are
faint, and have blue featureless spectra, which are typical of young
stars. The fact that there are only seemingly young stars very close
to the MBH, while there is a mixture of young and old stars farther
out raises a {}``Nature vs Nurture'' question. Is this an essentially
random variation in the stellar population, which can be explained
in terms of normal star formation processes ({}``Nature''), or is
this a result some systematic effects of the unique extreme environment
very near the MBH ({}``Nurture'')? It is interesting to note that
a cluster of blue stars exists also around the \( \sim \! 3\times 10^{7}\, M_{\odot } \)
MBH in the galaxy M31 (Lauer et al. \cite{Lau98}).

If these stars are indeed the products of their environment, then
there are two options to consider. First, this could be the result
of an unusual mechanism of star formation (Morris \cite{Mor93}),
in which case the stars are indeed young and dynamically unrelaxed,
and so do not convey direct information on the dynamical processes
in near the MBH. Second, this could be a results of unusual stellar
evolution, so that the stars only appear young, but are in fact old
and dynamically relaxed. This chapter will focus on the second possibility
(\S\ref{sec:Collider}). However, before we discuss possible mechanisms
for modifying stellar evolution, it is useful to review some results
from stellar dynamics theory.

\section{Stellar dynamics near a black hole}

\label{sec:Dynamics}

The stellar dynamical term {}``stellar collision'' is not limited
to the case of actual physical contact between stars, but refers to
any gravitational interaction where the stars exchange momentum or
energy. The dynamical processes in a gravitating stellar system can
be summarized by considering stellar collisions as function of their
distance scale. The reader is referred to Binney \& Tremaine (\cite{Bin87})
for a detailed overview.

On the largest scale, the motion of the star is determined by the
sum of interactions with all the other stars, that is, by the smooth
gravitational potential of the system. Two-body interactions occur
on a shorter length scale, when two stars approach each other to the
point where their mutual interaction dominates over that of the smoothed
potential. Two-body interactions randomize the stellar motions and
lead to the relaxation of the system. In the course of relaxation,
the stars, whose mass range spans 2--3 orders of magnitude, are driven
toward equipartition. However, equipartition cannot be achieved in
the presence of a central concentration of mass (in particular a central
MBH). When two stars, which are initially on the same orbit (and therefore
have the same velocity) interact, the massive one will slow down and
the lighter one will speed up. Since the radius of the orbit depends
only on the star's \emph{specific} energy, and not its total energy,
the massive star will sink to the center, while the lighter star will
drift outwards. Over time, this process leads to {}``mass segregation''---the
more massive stars are concentrated near the MBH and the lighter stars
are pushed out of the inner region.

Occasionally, two-body interactions will eject a star out of the system
altogether, thereby taking away positive energy from the system. The
system will then become more bound and compact, the collision rate
will increase, more stars will be ejected, and the result will be
a runaway process. This process is called the {}``gravothermal catastrophe'',
or {}``core collapse'', and is linked to the fact that self-gravitating
systems have a negative heat capacity---they become hotter when energy
is taken out. Core collapse, if unchecked, will lead to the formation
of an extremely dense stellar core surrounded by a diffuse extended
halo.

Once the density becomes high enough, very short range inelastic collisions
are no longer extremely rare, and the fact that the stars are not
point masses but have internal degrees of freedom starts to play a
role. In such collisions energy is extracted from the orbit and invested
in the work required to raise tides on the stars, or strip mass from
them. The tidal energy is eventually dissipated in the star and radiated
away. If the collision is slow, as it is in the core of a globular
cluster where there is no MBH, then the typical initial orbit is just
barely unbound. In this case, the tidal interaction may extract enough
orbital energy for {}``tidal capture'', and lead to the formation
of a tightly bound, or {}``hard'' binary (tight, because tidal forces
become effective only when the two stars are very close to each other).
Hard binaries are a heat source for the cluster and play a crucial
role in arresting core collapse. When a third star collides with a
hard binary, it will tend to gain energy from the binary, thereby
injecting positive energy to the cluster, while the binary becomes
harder still.

When the stars orbit a central MBH, the collisions are fast (The Keplerian
velocity near the MBH exceeds the escape velocity from the star) and
the initial orbits are very unbound (hyperbolic). Even very close
fly-bys cannot take enough energy from the orbit to bind the two stars,
and so they continue on their way separately after having extracted
energy and angular momentum from the orbit. The stars can radiate
away the excess heat on a time scale shorter than the mean time between
collisions, but it is harder to get rid of the excess angular momentum.
Magnetic breaking (the torque applied to a star when the stellar wind
resists being swept by the rotating stellar magnetic field), typically
operates on time scales similar to the stellar lifetime. It is therefore
likely that high rotation is the longest-lasting dynamical after effect
of a close hyperbolic encounter, and that stars in a high density
cusp are spun-up stochastically by repeated collisions (\S\ref{sec:spinup}).
Finally, at zero range, almost head-on stellar collisions can lead
to the stripping of stellar envelopes (\S\ref{sec:Cusp}), the destruction
of stars, or to mergers that result in the creation of {}``exotic
stars''. These are stars that cannot be formed in the course of normal
stellar evolution, such as a Thorne-Zytkow object, which is an accreting
neutron star embedded in a giant envelope (Thorne \& Zytkow \cite{Tho75}).

\subsection{Physical scales}

\label{sec:scales}

There are several important timescales and lengthscales that govern
the dynamics of the stellar system and MBH. They are listed here with
estimates of their value in the Galactic Center. A solar type star
and \( M_{\bullet }=3\times 10^{6}\, M_{\odot } \) (Genzel et al.
\cite{Gen00}) are assumed throughout. Physical lengths are expressed
also as angular sizes assuming that the distance to the Galactic Center
is \( R_{0}=8\, \mathrm{kpc} \) (Reid \cite{Rei93}).

\subsubsection{Timescales}

The dynamical time, or orbital time, \( t_{d} \), is the time it
takes a star to cross the system \begin{equation}
t_{d}\sim \frac{r}{v}\sim 2\pi \sqrt{\frac{\mathrm{r}^{3}}{GM_{\mathrm{tot}}}}\sim 2\times 10^{5}\, \mathrm{yr}\, (\mathrm{at}\, 3\, \mathrm{pc})\sim 300\, \mathrm{yr}\, (\mathrm{at}\, 0.03\, \mathrm{pc})\, ,
\end{equation}
 where \( r \) is the typical size of the system and \( M_{\mathrm{tot}} \)
is the total mass enclosed in radius \( r \).

The 2-body relaxation time, \( t_{r} \), is related to the 1D velocity
dispersion \( \sigma  \), the mean stellar mass \( \left\langle M_{\star }\right\rangle  \)
and the stellar number density \( n_{\star } \) by\begin{equation}
t_{r}\sim \frac{0.34\sigma ^{3}}{G^{2}\left\langle M_{\star }\right\rangle ^{2}n_{\star }\ln \Lambda }\sim 10^{9}\, \mathrm{yr}\, ,
\end{equation}
 where \( \log \Lambda  \) is the Coulomb logarithm, the logarithm
of the ratio between the largest and smallest impact parameters possible
in the system for elastic collisions. Because the relaxation timescale
in the Galactic Center is shorter than the age of the Galaxy (\( \sim 10 \)
Gyr), the old stars are expected to be well relaxed by now.

The mass segregation timescale is of the same order as the relaxation
timescale, \begin{equation}
t_{\mathrm{seg}}\sim t_{r}\, .
\end{equation}
 The rate (per star) of grazing collisions between two stars of mass
and radius \( M^{a}_{\star } \), \( R_{\star }^{a} \) and \( M_{\star }^{b} \),
\( R_{\star }^{a} \), each, is \begin{equation}
\label{eq:tcoll}
t^{-1}_{c}=4\sqrt{\pi }n_{\star }\sigma \left( R^{a}_{\star }+R^{b}_{\star }\right) ^{2}\left[ 1+\frac{G\left( M^{a}_{\star }+M^{b}_{\star }\right) }{2\sigma ^{2}\left( R^{a}_{\star }+R^{b}_{\star }\right) }\right] \sim 10^{-9}\, \mathrm{yr}^{-1}\, (\mathrm{at}\, 0.03\, \mathrm{pc})\, ,
\end{equation}
 where it is assumed that the stars follow a mass independent Maxwell-Boltzmann
velocity distribution with velocity dispersion \( \sigma  \) (this
is a good approximation near the MBH, see \S\ref{sec:relax}). There
are two contributions to the total rate, one due to the geometric
cross-section (first term in the square brackets) and one due to {}``gravitational
focusing'' (second term in the square brackets). Gravitational focusing
expresses the fact that the two stars do not move on straight lines,
but are attracted to each other. This effect is important when the
typical stellar velocities are much smaller than the escape velocity
from the stars, \( \sigma ^{2}<GM_{\star }/2R_{\star }=v_{\mathrm{esc}}^{2}/4 \).

\subsubsection{Lengthscales}

The size of the event horizon of a non-rotating black hole, the Schwarzschild
radius, is \begin{equation}
r_{s}=\frac{2GM_{\bullet }}{c^{2}}=9\times 10^{11}\, \mathrm{cm}\sim 3\times 10^{-7}\, \mathrm{pc}\sim 8\, \mu \mathrm{arcsec}\, .
\end{equation}

The tidal radius, \( r_{t} \), is the minimal distance from the MBH
where the stellar self-gravity can still resist the tidal forces of
the MBH. If the star's orbit takes it inside the tidal radius, it
will be disrupted, and roughly half of its mass will fall into the
MBH, while the other half will be ejected (e.g. Ayal, Livio \& Piran
\cite{Aya00}). The exact value of the tidal radius depends on the
stellar structure and the nature of the orbit, and up to a factor
of order unity is given by \begin{equation}
r_{t}\sim R_{\star }\left( \frac{M_{\bullet }}{M_{\star }}\right) ^{1/3}=10^{13}\, \mathrm{cm}\sim 3\times 10^{-6}\, \mathrm{pc}\sim 80\, \mu \mathrm{arcsec}\, .
\end{equation}
 Tidal disruption is relevant as long as the tidal radius lies outside
the event horizon. Since \( r_{t}\propto M_{\bullet }^{1/3} \) ,
while \( r_{s}\propto M_{\bullet } \), there exists a maximal MBH
mass for tidal disruption, which for a solar type stars is \( \sim \! 10^{8}\, M_{\odot } \).

The radius of influence, \( r_{h} \), is the region where the MBH
potential dominates the dynamics. If the MBH is embedded in an isothermal
stellar system (i.e. \( \sigma  \) is constant), then the radius
of influence can be defined as \begin{equation}
r_{h}=\frac{GM_{\bullet }}{\sigma ^{2}}\sim 10^{19}\, \mathrm{cm}\sim 3\, \mathrm{pc}\sim 80\, \mathrm{arcsec}\, .
\end{equation}
 In practice, the distribution is not isothermal and \( \sigma  \)
is not constant, and so \( r_{h} \) is evaluated loosely by choosing
a representative value of \( \sigma  \) far enough from the MBH.
The stellar mass enclosed within \( r_{h} \) is of the same order
as the mass of the MBH.

\subsection{A relaxed stellar system around a MBH}

\label{sec:relax}

The relaxed, quasi steady-state density distribution of a single-mass
stellar population around a MBH is (Bahcall \& Wolf \cite{Bah76};
see also Binney \& Tremaine \cite{Bin87} for a simple derivation)

\begin{equation}
n_{\star }\propto r^{-7/4}\, .
\end{equation}
 When the the stellar population consists of a spectrum of masses,
\( M_{1}<M_{\star }<M_{2}, \) the stellar distribution function (DF)
very near the MBH has the form (Bahcall \& Wolf\cite{Bah77}) \begin{equation}
f_{M}(\epsilon )\propto \epsilon ^{p_{M}}\, ,\qquad n_{\star }\propto r^{-3/2-p_{M}}\, ,\qquad p_{M}\equiv \frac{M}{4M_{2}}\, ,
\end{equation}
 where \( -\epsilon  \) is the total specific energy of the star
and \( f_{M}\equiv 0 \) for \( \epsilon <0 \). The velocity dispersion
of this DF (see equation \ref{eq:Jeans} below) is almost independent
of the stellar mass,

\begin{equation}
\sigma _{M}^{2}=\left( \frac{1}{5/2+p_{M}}\right) \frac{GM_{\bullet }}{r}\, ,
\end{equation}
 which implies that \( \sigma _{M}^{2} \) changes by less than 10\%
over the entire mass range, in marked contrast to the \( \sigma _{M}^{2}\propto M^{-1}_{\star } \)
dependence of equipartition. This result justifies the approximation
that the velocity dispersion in a relaxed stellar system around a
MBH is mass-independent.

The Bahcall-Wolf solution applies to point particles. This assumption
no longer holds very near the MBH, where the collision rate is high
because of the very high stellar density. Stars on tight orbits around
the MBH cannot survive for long, and so eventually most of the population
there will consist of stars that are on very wide, marginally bound
(parabolic) orbits, which spend only a small fraction of their time
in the collisionally dominated region. These marginally bound stars
have a flatter spatial distribution, of the form (e.g. Binney \& Tremaine
\cite{Bin87}, p. 551)

\begin{equation}
n_{\star }\propto r^{-1/2}\, .
\end{equation}

\begin{figure}[tp]
{\centering \vspace*{1mm}\resizebox*{0.4\textwidth}{!}{\includegraphics{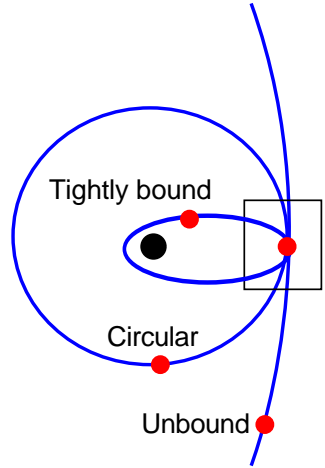}} \par}

\caption{A schematic representation of the different types of stellar orbits
that can be observed in a small region near the MBH.}

\label{fig:Orbits}
\end{figure}

The stars in any volume element near the MBH have a distribution of
orbits (Figure \ref{fig:Orbits}): some are more bound than circular
(i.e. their velocity is smaller than the circular velocity \( v_{c} \)),
some are less bound than circular, some are unbound to the MBH (but
bound by the total mass of the MBH and stars) . The distribution of
orbits is directly tied to the spatial distribution through the Jeans
Equation,

\begin{equation}
\label{eq:Jeans}
\frac{GM_{\bullet }}{r\sigma ^{2}}=\frac{v^{2}_{c}}{\sigma ^{2}}=-\frac{d\ln n_{\star }}{d\ln r}-\frac{d\ln \sigma ^{2}}{d\ln r}\, .
\end{equation}
 The Jeans equation is essentially a re-statement of the continuity
equation of the stellar orbits in phase space in terms of averaged
quantities, the mean stellar density and velocity dispersion. Here
it is given for the simplest case of a steady state, isotropic, non-rotating
system. The steady-state assumption is justified because the dynamical
timescale is much shorter than the relaxation timescale. The assumptions
of isotropy and non-rotation are observationally justified.

\begin{figure}[tp]
{\centering \vspace*{1mm}\resizebox*{0.7\textwidth}{!}{\includegraphics{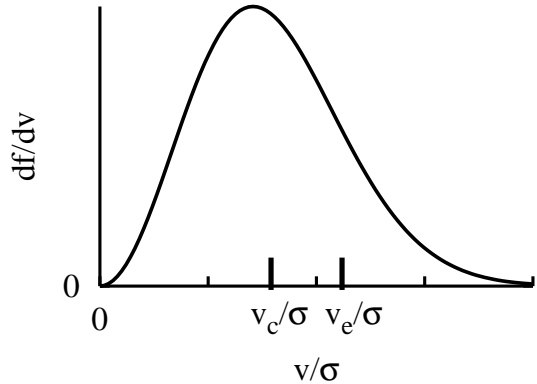}} \par}

\caption{The fraction of orbits in the Maxwell-Boltzmann distribution as function
of the normalized velocity \protect\( v/\sigma \protect \). The circular
velocity \protect\( v_{c}\protect \) and the escape velocity \protect\( v_{e}=\sqrt{2}v_{c}\protect \)
are marked for a density distribution with \protect\( \alpha =1.5\protect \).
The region \protect\( v>v_{e}\protect \) lies under the exponential
tail of the DF, and so the fraction of stars with unbound orbits is
a strongly decreasing function of \protect\( \alpha \protect \). }

\label{fig:MB}
\end{figure}

Very near the MBH the velocity dispersion is Keplerian, \( \sigma ^{2}\propto r^{-1} \),
and so for any power-law cusp \( n_{\star }\propto r^{-\alpha } \)
the Jeans equation implies that \begin{equation}
\label{eq:sign}
\frac{v^{2}_{c}}{\sigma ^{2}}=\alpha +1\, .
\end{equation}
 The steeper the cusp (larger \( \alpha  \)), the larger the ratio
between \( v_{c} \) and \( \sigma  \), and so the fraction of loosely
bound stars or unbound stars is smaller (Figure \ref{fig:MB}). Because
unbound stars have wide orbits and spend most of their time far away
from the MBH, the stellar population in a shallow cusp is well mixed
and representative of the average population over a large volume.
In contrast, the stellar population in a steep cusp is localized and
can therefore develop and maintain properties that differ from those
of the general population.

\section{The stellar collider in the Galactic Center}

\label{sec:Collider}

The potential for probing a new regime of stellar dynamics near the
MBH in the Galactic Center is best illustrated by comparing the collisional
timescale there with that in the cores of the densest globular clusters,
which for long served as laboratories for the study of collisional
processes. In a dense globular cluster, \( n_{\star }\! \sim \! 10^{6}\, M_{\odot }\, \mathrm{pc}^{-3} \)
and \( \sigma \! \sim \! 10\, \mathrm{km}\, \mathrm{s}^{-1} \), whereas
in the Galactic Center, the density may be as high as \( n_{\star }\! \sim \! 10^{8}\, M_{\odot }\, \mathrm{pc}^{-3} \)
(\S\ref{sec:Cusp}) and \( \sigma \! \sim \! 1000\, \mathrm{km}\, \mathrm{s}^{-1} \).
The timescale for collisions between solar type stars in a globular
cluster can be estimated from equation (\ref{eq:tcoll}) to be almost
\( 10^{10}\, \mathrm{yr} \), roughly the age of the Galaxy and of
a solar type star, whereas it is only \( \sim \! 5\times 10^{8}\, \mathrm{yr} \)
in the inner 0.02 pc of the Galactic Center. These estimates imply
that while physical collisions are only marginally relevant in the
cores of the densest globular clusters, they completely dominate the
dynamics in the innermost part of the MBH cusp%
\footnote{The probability for \emph{avoiding} a collision over a time \( t \)
is \( \exp \left( -t/t_{c}\right)  \). 
}.

\subsection{The case for a dense stellar cusp in the Galactic Center}

\label{sec:Cusp}

Theoretical expectations lead us to expect a relaxed stellar cusp
around the MBH in the Galactic Center. Does such a cusp indeed exist
there? The answer depends critically on the problem of identifying
which of the observed stars are dynamically relaxed, since only those
faithfully trace the underlying old stellar population. The analysis
presented here shows that it is possible to interpret the available
observations \emph{self-consistently} in the framework of a high density
cusp. However, the reader should keep in mind that the issue is an
empirical one, and as such may be subject to revisions when more and
better data is obtained about the stars near the MBH.

\begin{figure}[tp]
{\centering \vspace*{1mm}\resizebox*{1\textwidth}{!}{\includegraphics{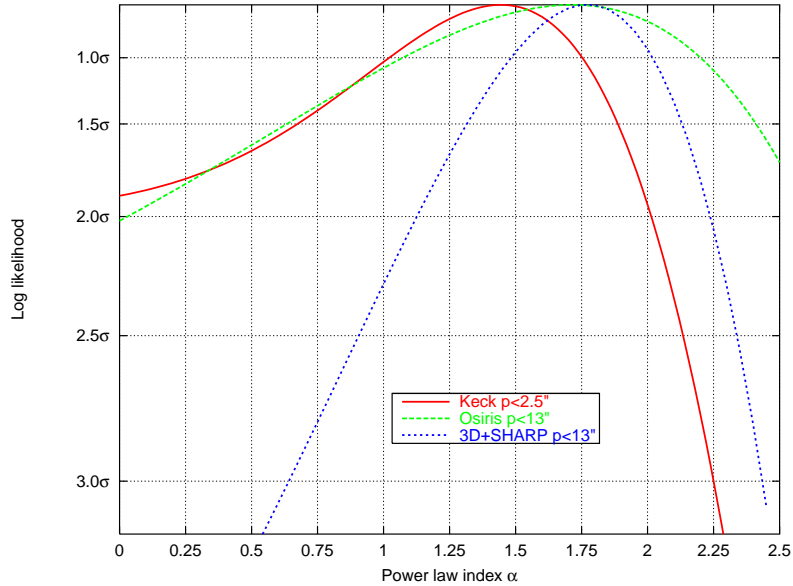}} \par}

\caption{A Maximum likelihood analysis of the surface density distribution
of stars near SgrA\protect\( ^{\star }\protect \) for a 3D stellar
density distribution \protect\( n_{\star }\propto r^{-\alpha }\protect \)
(Alexander \cite{Ale99a}). Three different data sets (Blum et al.
\cite{Blu96}; Genzel et al. \cite{Gen96}; Eckart \& Genzel \cite{Eck97};
Ghez et al. \cite{Ghe98}) indicate that the most likely value for
\protect\( \alpha \protect \) lies in the range \protect\( \sim \! 3/2\protect \)
to \protect\( \sim \! 7/4\protect \), which is the theoretically
predicted range for a relaxed stellar system around a MBH (Bahcall
\& Wolf \cite{Bah77}). Order of magnitude estimates (\S\ref{sec:scales})
suggest that the stellar system around the MBH in the Galactic Center
has undergone two-body relaxation. (Reprinted with permission from
\emph{The Astrophysical Journal}).\label{fig:ML_pl}}
\end{figure}

Direct evidence for the existence of a cusp comes from the analysis
of star maps, which show a concentration of stars toward the center.
Assuming a 3D density distribution of the form \( n_{\star }\propto r^{-\alpha } \),
the corresponding projected 2D surface density can be compared to
the observed distribution to find the most likely value of \( \alpha  \).
Figure \ref{fig:ML_pl} shows the likelihood curves for \( \alpha  \)
based on three independent star maps, after all the stars that were
spectroscopically identified as young were taken out of the sample
(the faint blue stars nearest to SgrA\( ^{\star } \) are included
only in the Keck data set, but not in the other two). The most likely
value for the density power-low index \( \alpha  \) lies in the range
\( \sim 1.5 \)--\( 1.75 \). A flat core (\( \alpha \! \sim \! 0 \)),
such as exists in globular clusters, is decisively rejected. Similarly,
a likelihood test for the maximal size of a flat inner core indicates
that such a core, if it exists, is be smaller than \( \sim \! 0.1 \)
pc (\( 2.5'' \)). It can be shown that extinction by interstellar
dust is unlikely to bias these results by a significant amount.

\begin{figure}[tp]
{\centering \vspace*{1mm}\resizebox*{1\textwidth}{!}{\rotatebox{270}{\includegraphics{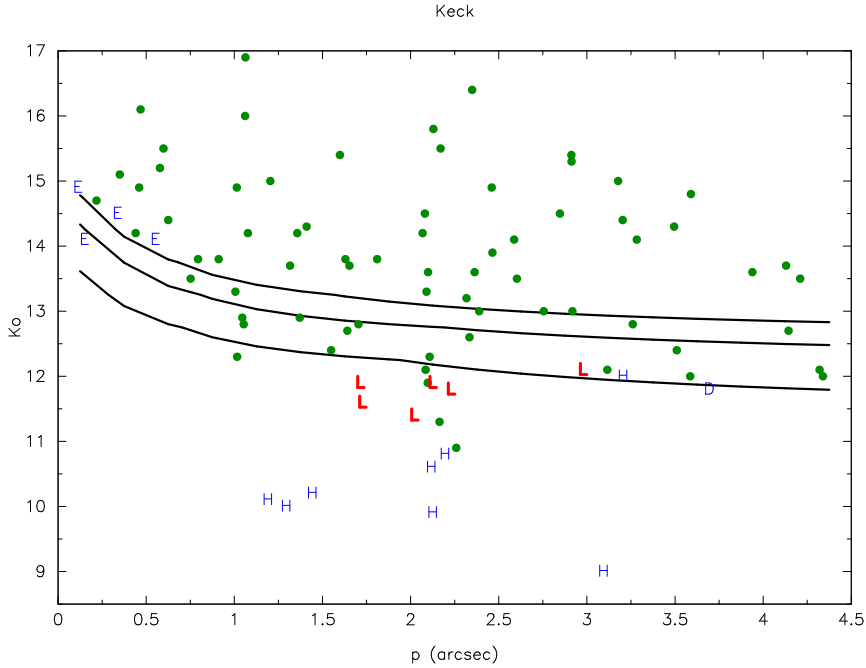}}} \par}

\caption{Evidence for collisional destruction of bright giant envelopes in
a high density stellar cusp around the MBH in the Galactic Center
(Alexander \cite{Ale99a}). The apparent stellar \protect\( K\protect \)-band
magnitude is plotted against the projected angular distance from the
black hole, \protect\( p\protect \) (Keck data from Ghez et al. \cite{Ghe98}).
The ages of the stars marked by circles are unknown, but it is likely
that most of them are old, and therefore dynamically relaxed. Stars
marked by {}``L'' are spectroscopically identified as old stars.
Stars marked by {}``H'' are spectroscopically identified as young
stars and are not dynamically relaxed. Such stars are not expected
to be affected by collisions because of their short lifetimes. The
stars marked by {}``E'' have featureless blue spectra and are either
young stars or old stars that were affected by the extreme conditions
very near the black hole. The three contour lines represent detailed
model predictions for the decrease in surface density of bright stars
due to collisional destruction in a high density \protect\( n_{\star }\propto r^{-3/2}\protect \)
stellar cusp. The stellar density reaches a value of \protect\( \sim 4\times 10^{8}\, \textrm{M}_{\odot }\, \mathrm{pc}^{-3}\protect \)
at \protect\( r=0.25''\protect \) (0.01 pc), which is 9 orders of
magnitude higher than in the Solar Neighborhood, and almost 3 orders
of magnitude higher than in the densest globular cluster core. The
model predicts, on average, 1.5 (top contour), 1.0 (central contour),
and 0.5 (bottom contour) dynamically relaxed stars per 0.25 arcsecond
bin that are brighter than the contour level. This is consistent with
the observed trend in the surface density distribution. (Reprinted
with permission from \emph{The Astrophysical Journal}).}

\label{fig:SGdepletion}
\end{figure}

Additional evidence for the existence of a very high density cusp
comes from the observed gradual depletion of the luminous giants toward
the MBH in the inner 0.1 pc (Figure \ref{fig:SGdepletion}). Luminous
red giants have very large extended envelopes, and therefore a large
cross-section for collisions with other stars. When the impact parameter
is a small fraction of the giant's radius, the envelope may be stripped,
leaving behind an almost bare burning core. This will make the star
effectively invisible in the infrared (IR) because the IR spectral
range lies in the Raleigh-Jeans part of the stellar blackbody spectrum,
and so the IR luminosity scales as \( L_{\mathrm{IR}}\propto R_{\star }^{2}T_{\mathrm{eff}} \)
while the total luminosity scales as \( L_{\star }\propto R_{\star }^{2}T_{\mathrm{eff}}^{4} \).
Suppose that the collision disperses the envelope of a \( \sim \! 100\, R_{\odot } \)
red supergiant and leaves a \( \sim \! 1\, R_{\odot } \) burning
core. In order to maintain the total stellar luminosity, the effective
temperature will have to rise by a factor 10, which will result in
a decrease of the IR luminosity by a factor of 1000 (7.5 magnitudes).\\

Figure \ref{fig:SGdepletion} compares a theoretical prediction for
the collisional depletion of luminous giants with the data. The match
with the observed trend is remarkably good, given the fact that no
attempt was made to fit the data. The calculation is based on detailed
modeling of expected numbers, sizes, luminosities and lifetimes of
giants in the population, on cross-sections for envelope disruption
that were calibrated by hydrodynamical simulations, and on a stellar
density cusp that is normalized by dynamical estimates of the enclosed
mass.

It should be noted that the total mass loss rate from these collisions
is smaller than that supplied by the strong stellar winds of the blue
supergiants in the inner few arcseconds, and so stellar collisions
are not a dominant source of mass supply to the MBH at this time.

The self-consistent picture that emerges from this analysis is that
the stars near the MBH in the Galactic Center, which are expected
to be dynamically relaxed, are indeed concentrated in a stellar cusp
of the form predicted by theory for a relaxed system. The very high
stellar density in the inner few 0.01 pc leads to frequent collisions
that destroy the envelopes of giant stars, thereby explaining the
gradual depletion in the number of luminous giants toward the center.
The central cluster of faint blue stars in the inner \( 0.5'' \)
coincides with the collisionally dominated region. It is therefore
relevant to consider dynamical explanations for their nature and appearance
as an alternative to assuming that they are newly formed, unrelaxed
stars. The concentration of such a distinct population in a small
volume is consistent with the tightly bound orbits that are typical
of a steep cusp (\S\ref{sec:relax}).

\subsection{Tidal spin-up}

\label{sec:spinup}

It is inevitable that in a system where the stellar density is high
enough for collisional destruction of giants, smaller stars that escape
destruction will still suffer very close encounters. As described
above (\S\ref{sec:Dynamics}), usually such collisions cannot bind
the two stars, and the longest lasting after effect, apart from possible
mass loss, is fast rotation. Fast rotation and mass loss have the
potential to affect stellar evolution and modify the appearance of
the stars (see discussion in Alexander \& Kumar \cite{Ale01b}; Alexander
\& Livio \cite{Ale01c}). Although detailed predictions of the observational
consequences are still not available, it is of interest to estimate
the magnitude of the spin-up effect.

When the tidal deformations in the star are small, the change in the
angular velocity of a star of mass \( M_{\star } \) and radius \( R_{\star } \)
due to an encounter with a mass \( M \) can be described by a linear
multipole expansion in the periapse distance \( r_{p} \) (distance
of closest approach) by (Press \& Teukolsky \cite{Pre77})

\begin{equation}
\label{eq:dW}
\Delta \widetilde{\Omega }=\frac{\widetilde{M}^{2}}{\widetilde{I}\widetilde{v}_{p}}\sum ^{\infty }_{l=2}\frac{T_{l}\left( \eta ,e\right) }{\widetilde{r}^{2l+1}_{p}}\, ,
\end{equation}
 where the tilde symbol denotes quantities measured in units of \( G=M_{\star }=R_{\star }=1 \),
and rigid body rotation is assumed. \( \widetilde{v}_{p} \) is the
relative velocity at periapse, \( \widetilde{I} \) is the star' s
moment of inertia, and \( T_{l} \) the tidal coupling coefficient
of the \( l \)'th moment. In these units, \( \widetilde{\Omega }=1 \)
is the centrifugal breakup angular velocity, where the star sheds
mass from its equator. The tidal coupling coefficients depend on the
star's structure and on the orbital parameters through the quantity
\( \eta =\widetilde{r}^{3/2}_{p}\left/ \sqrt{1+\widetilde{M}}\right.  \)
and the orbital eccentricity \( e \). The tidal coefficient \( T_{l} \)
can be calculated numerically for any given stellar model and orbit.

The formal divergence of \( \Delta \widetilde{\Omega } \) as \( \widetilde{r}_{p} \)
decreases indicates that most of the contribution comes from close
collisions, where the linear analysis breaks down. The non-linear
processes, which truncate the divergence, have to be investigated
by hydrodynamical simulations (see \S\ref{sec:Scatter}). These reveal
that as \( \widetilde{r}_{p} \) decreases towards 1, \( \Delta \widetilde{\Omega } \)
first increases faster than predicted by the linear analysis, but
then it reaches a maximal value at the onset of mass loss, since the
ejecta carry away the extra angular momentum.

\begin{figure}[tp]
{\centering \vspace*{1mm}\resizebox*{1\textwidth}{!}{\rotatebox{270}{\includegraphics{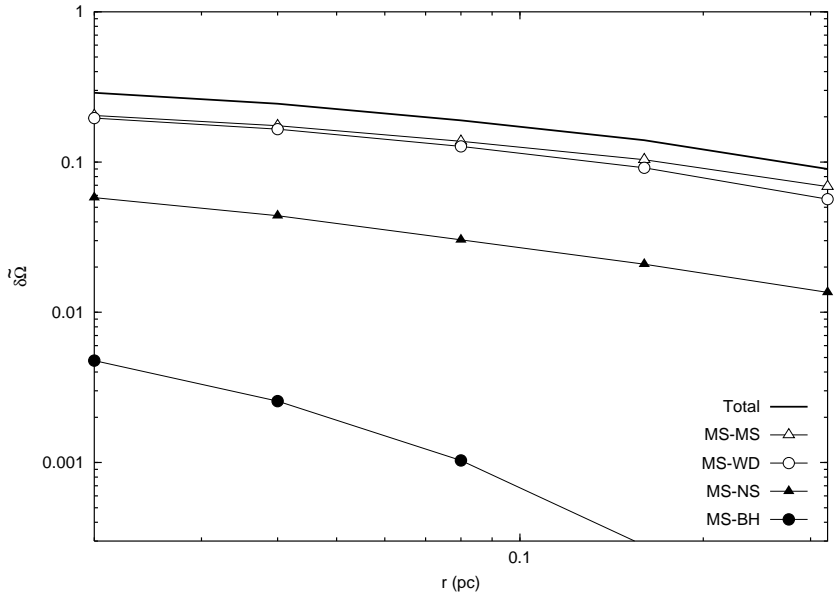}}} \par}

\caption{The average spin-up of a solar type star by star-star tidal interactions
over 10 Gyr as function of distance from the black hole in the Galactic
Center (Alexander \& Kumar \cite{Ale01b}). An \protect\( \alpha =1.5\protect \)
density cusp is assumed. The rotation grows over time in a random
walk fashion by repeated close passages. \protect\( \delta \widetilde{\Omega }=1\protect \)
corresponds to rotation at the centrifugal break-up velocity. In addition
to the total spin-up, the separate contributions from collisions with
main sequence stars (MS), white dwarfs (WD), neutron stars (NS) and
stellar black holes (BH) are shown. (Reprinted with permission from
\emph{The Astrophysical Journal}).}

\label{fig:dw_r}
\end{figure}

Over its lifetime, a star will undergo many tidal encounters, randomly
orientated relative to its spin axis, and will be spun-up in a random
walk manner. The cumulative effect can be large. Figure \ref{fig:dw_r}
shows the predicted average spin-up of a solar type stars over 10
Gyr in the Galactic Center as function of distance from the MBH (Alexander
\& Kumar \cite{Ale01b}). The calculation assumes an \( \alpha =1.5 \)
density cusp, a model for the distribution of stellar masses in the
population, and inefficient magnetic breaking. On average, solar type
stars in a large volume around the black hole are spun-up to 10\%--30\%
of the break-up angular velocity, or 20 to 60 times faster than is
typical in the field. The effect falls off only slowly with distance
because the higher efficiency of tidal interactions in slower collisions
far from the black hole offsets the lower collision rate there.

\subsection{Tidal scattering}

\label{sec:Scatter}

Tidal scattering is another mechanism that can affect the internal
structure of a significant fraction of the stars around the MBH. Unlike
the tidal spin-up process discussed in \S\ref{sec:spinup}, tidal
scattering does not require a very high stellar density, since it
is driven by the global response of the system to the existence of
a mass sink, the MBH, in its center.

Some of the mass that feeds the growth of a MBH in a galactic center
is supplied by tidal disruption of stars that are scattered into low
angular momentum orbits ({}``loss-cone'' orbits). When the MBH mass
is small enough so that the tidal radius is larger than the event
horizon, \( r_{t}>r_{\mathrm{s}} \), the star is tidally disrupted
before crossing the event horizon. The accretion of stellar debris
from such events may give rise to observable {}``tidal flares''
(Frank \& Rees \cite{Fra76}). Significant theoretical efforts have
gone into estimating the rates, timescales, luminosities and spectra
of the flares (e.g. Ulmer, Paczy\'{n}ski, \& Goodman, \cite{Ulm98};
Magorrian \& Tremaine \cite{Mag99}; Ayal, Livio \& Piran \cite{Aya00}),
in the hope that they can be used to detect MBHs in the centers of
galaxies. There is today only marginal evidence for the detection
of such flares (e.g. Renzini et al. \cite{Ren95}; Komossa \& Bade
\cite{Kom99a}; Komossa \& Greiner \cite{Kom99b}).

\begin{figure}[tp]
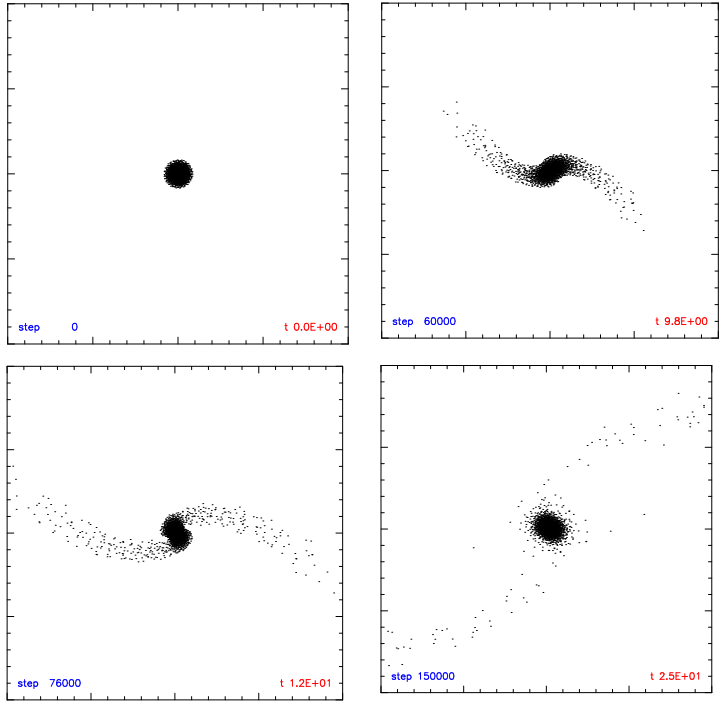

{\centering \vspace*{1mm}\par}

{\centering \begin{tabular}{cc}
\resizebox*{0.4\textwidth}{!}{\rotatebox{270}{\includegraphics{bh_tide_1.ps}}} &
\resizebox*{0.4\textwidth}{!}{\rotatebox{270}{\includegraphics{bh_tide_121.ps}}} \\
\resizebox*{0.4\textwidth}{!}{\rotatebox{270}{\includegraphics{bh_tide_153.ps}}} &
\resizebox*{0.4\textwidth}{!}{\rotatebox{270}{\includegraphics{bh_tide_301.ps}}} \\
\end{tabular}\par}

\caption{Snapshots from a Smoothed Particle Hydrodynamics (SPH) simulation
of a star undergoing an extreme non-disruptive tidal interaction ({}``tidal
scattering'') as it passes near a massive black hole. Time is measured
in units of the star's dynamical timescale. The star passes near the
black hole (located outside of the frame) on a parabolic orbit with
a peri-distance 1.5 times larger than the tidal disruption distance.
Shortly after periapse passage (\protect\( t=12\protect \)) the star
appears to be on the verge of breaking in two. However, by the end
of the simulation, the two fragments coalesce, leaving a distorted,
mixed and rapidly rotating bound object. }

\label{fig:BHtide}
\end{figure}

The effect of MBH's tidal field is not limited to tidal disruption.
For every star that is actually disrupted, there are stars with \( r_{p}\gtrsim r_{t} \)
that narrowly escape tidal disruption by the central BH after being
subjected to extreme tidal distortion, spin-up, mixing and mass-loss,
which may affect their evolution and appearance (Alexander \& Livio
\cite{Ale01c}). Figure \ref{fig:BHtide} shows Smoothed Particle
Hydrodynamics%
\footnote{SPH is an algorithm for simulating the hydrodynamics of 3D self-gravitating
fluids, which is commonly used in the study of stellar collisions
(Monaghan \cite{Mon92}). The star is represented by discrete mass
elements, each distributed smoothly over a small sphere so that the
density peaks in the center and falls to zero at the edge. The total
density at a point is the sum of densities in all overlapping spheres
that include the point. The resulting density field is continuous
and differentiable, and so its thermodynamic properties can be evaluated
everywhere once an equation of state is specified. Every time step,
the positions of the mass elements are updated according to the gravitational
force and the pressure gradient, and the sphere sizes are readjusted
to reflect the changes in the local density. 
} (SPH) simulation of a star passing by an MBH just outside the tidal
disruption radius. To leading order, the effects of tidal scattering
are a function of the penetration parameter \( \beta =r_{t}/r_{p} \)
only, and are independent of the MBH mass,

\begin{equation}
\Delta \widetilde{\Omega }\simeq \frac{T_{2}\left( \beta ^{-3/2}\right) }{\sqrt{2}\widetilde{I}}\beta ^{9/2}\, ,
\end{equation}
 which follows from equation (\ref{eq:dW}) for a parabolic orbit
and for \( M_{\bullet }\gg M_{\star } \). As will be argued below,
a large fraction of these {}``tidally scattered'' stars survive
eventual orbital decay and disruption, and so remain in the system
as relics of the epoch of tidal processes even after the MBH becomes
too massive for tidal disruption.

Dynamical analyses of the scattering of stars into the loss-cone orbits
(Lightman \& Shapiro \cite{Lig77}; Magorrian \& Tremaine \cite{Mag99})
show that tidally disrupted stars in galactic nuclei are typically
on slightly unbound orbits relative to the MBH and that they are predominantly
scattered into the loss-cone from orbits at the radius of influence
of the BH, \( r_{h} \). The scattering operates on a timescale that
is shorter than the dynamical timescale, and so the stars are scattered
in and out of the loss-cone several times during one orbital period.
Because of gravitational focusing, the cross-section for scattering
into a hyperbolic orbit with periapse \( \leq r_{p} \) scales as
\( r_{p} \), and not as \( r^{2}_{p} \) (Hills \cite{Hil75}; Frank
\cite{Fra78}), and so the number of stars with \( r_{t}\leq r_{p}\leq 2r_{t} \)
equals the number of stars that were disrupted by the MBH.

Tidal disruption is an important source of mass for a low-mass MBH
that accretes from a low-density galactic nuclear core, where mass
loss from stellar collisions is small (e.g. Murphy, Cohn \& Durisen
\cite{Mur91}). For the MBH in the Galactic Center, the total mass
in disrupted stars can be \( 0.25M_{\bullet } \) or even higher (Freitag
\& Benz 2001, in preparation). Since the enclosed stellar mass within
\( r_{h} \) is also \( \sim \! M_{\bullet } \), the tidally scattered
stars comprise a significantly high fraction of the stellar population
within the radius of influence of the MBH.

After the first periapse passage, the tidally scattered star will
be on a very eccentric orbit with a maximal radius (apoapse) of \( \lesssim 2r_{h} \).
Since the two body interactions that scattered it into the eccentric
orbit operate on a timescale that is shorter than the orbital period,
there is a significant chance that the star will be scattered again
off the orbit and miss the MBH. The chance of this happening is further
increased by the Brownian motion of the MBH relative to the dynamical
center of the stellar system. The amplitude of the Brownian motion
is much larger than the tidal radius, and it proceeds on the dynamical
timescale of the core (Bahcall \& Wolf \cite{Bah76}), which is comparable
to the orbital period of the tidally disturbed stars. The orbits of
the tidally scattered stars take them outside of \( r_{h} \), where
they are no longer affected by the relative shift between the BH and
the stellar mass. Therefore, on re-entry into the volume of influence,
their orbit will not bring them to the same periapse distance from
the MBH. Both the random motion of the MBH and the scattering off
the loss-cone by two-body interactions are expected to increase the
survival fraction to a significantly high value. More detailed calculations,
which integrate over the orbital distribution, are required to confirm
these qualitative arguments.

Rough estimates (Alexander \& Livio \cite{Ale01c}) indicate that
the Galactic Center may harbor \( 10^{4-5} \) tidally scattered stars.
These stars are expected to be on highly eccentric orbits, and so
there may be observable correlations between high orbital eccentricity
and the stellar properties.

\begin{figure}[tp]
{\centering \vspace*{1mm}\resizebox*{0.9\textwidth}{!}{\includegraphics{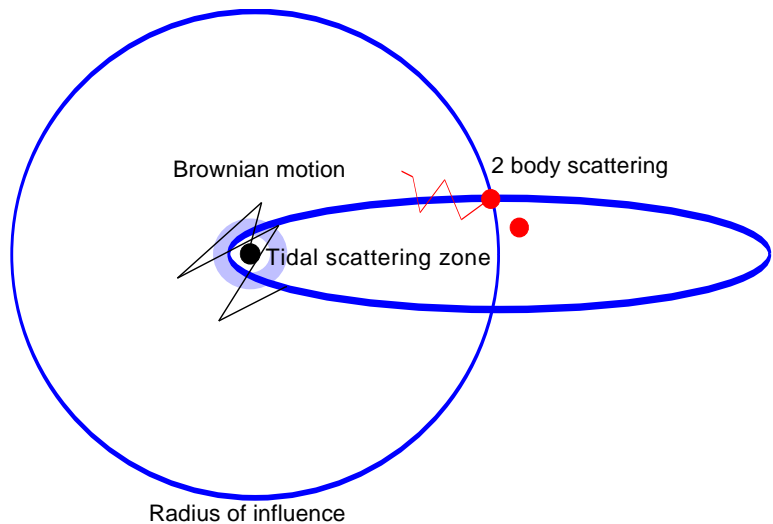}} \par}

\caption{A schematic representation of the tidal scattering process. A star
initially orbiting the MBH at the radius of influence is scattered
by a two body encounter into an extremely eccentric orbit that brings
it to the tidal scattering zone just outside the tidal disruption
radius. The star suffers an extreme, non-disruptive tidal interaction
with the MBH, and continues on its way out of the radius of influence,
where it is scattered by frequent two-body encounters. In the meanwhile,
the Brownian motion of the MBH due to its interactions with the stellar
system causes it to move away from its original position. Both these
random processes significantly increase the chances of the tidally
disturbed star to survive total disruption during subsequent orbits. }

\label{fig:TidalScatter}
\end{figure}

\section{The gravitational telescope in the Galactic Center}

\label{sec:Telescope}

The MBH in the Galactic Center is a telescope with a lens of effective
diameter \( \sim \! 4\times 10^{17}\, \mathrm{cm} \) (for a source
at infinity) and a focal length of \( \sim \! 2.5\times 10^{22}\, \mathrm{cm} \).
Unfortunately, Nature did not design it as an ideal telescope. A point
mass lens does not produce faithful images of the lensed sources,
the optical axis is heavily obscured by interstellar dust, and the
telescope points in a fixed direction, which is not of our choosing.
In fact, various estimates suggest that there are not enough luminous
sources in that direction for gravitational lensing to be important
for present day observations, although future, deep observations may
pick up lensing events (Wardle \& Yusef-Zadeh \cite{War92}; Alexander
\& Sternberg \cite{Ale99b}; Alexander \& Loeb \cite{Ale01d}). Nevertheless,
it is worthwhile to consider the possible roles of gravitational lensing
in the observations and study of the Galactic Center. This is important
not only in anticipation of future observations, but also because
the estimates of the lensing probability are quite uncertain (they
involve models of the unobserved far side of the Galaxy), and because
there are hints that lensing may not be quite as rare as predicted
(\S\ref{sec:Pinpoint}).

Gravitational lensing may be used to probe the dark mass (is it really
a MBH?) and the stars around it, and to locate the MBH on the IR grid,
where the stars are observed. On the other hand, gravitational lensing
can also complicate the interpretation of the observations since it
affects many of the observed properties of the sources: flux, variability,
apparent motion and surface density. IR flares due to lensing can
be confused with those due to fluctuations in the accretion flow,
and lensed images of background sources far behind the MBH can be
confused with stars that are truly near the MBH. This section will
focus on aspects of gravitational lensing that are, or may be relevant
for the Galactic Center. The reader is referred to Schneider, Ehlers
\& Falco (\cite{Sch92}) for a comprehensive treatment of the subject.

\subsection{Gravitational lensing by a point mass}

\begin{figure}[tp]
{\centering \vspace*{1mm}\resizebox*{0.9\textwidth}{!}{\includegraphics{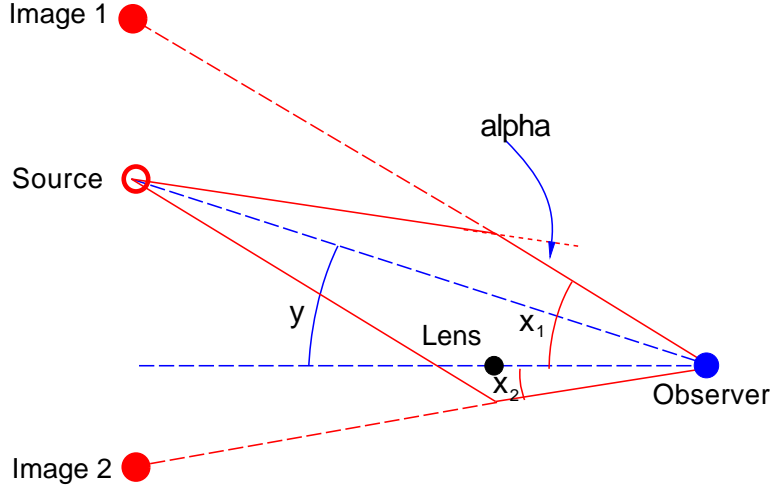}} \par}

\caption{The light ray diagram for lensing by a point mass. The light rays
from a source, at angular position \protect\( y\protect \) relative
to the observer--lens optical axis, are bent by the gravitational
lens and reach the observer from angles \protect\( x_{1}\protect \)
and \protect\( x_{2}\protect \), thereby appearing as two images.
Unlike a glass lens, the light bending angle \protect\( \alpha \protect \)
of a gravitational lens is inversely proportional to the impact parameter
to the lens (Equation \ref{eq:bend}). }

\label{fig:LightRays}
\end{figure}

To first and good approximation the lensing properties of the mass
distribution in the Galactic Center can be described as those of a
point mass, the MBH. Figure \ref{fig:LightRays} shows the light ray
diagram of lensing by a point mass in the small angle limit. The bending
angle is given by

\begin{equation}
\label{eq:bend}
\alpha =\frac{4GM_{\bullet }}{c^{2}b}\, ,
\end{equation}
 where \( b \) is the impact parameter of the light ray with respect
to the lens. Note that unlike a glass lens, where the bending angle
is zero when the ray goes through the lens center and increases with
the impact parameter, the bending angle of a gravitational lens diverges
towards the center and decreases with the impact parameter. It is
therefore not surprising that a gravitational lens does not produce
a faithful image of the lensed source, but rather breaks, warps and/or
flips the image. A point lens creates two images of the source, one
on either side of the lens. There are always two images in focus at
the observer, regardless of the distance of the source behind the
lens. The two images, the lens and the (unobserved) source all lie
on one line. The typical angular cross-section of the lens is given
by the Einstein angle,

\begin{equation}
\label{eq:ThetaE}
\theta ^{2}_{E}=\frac{4GM_{\bullet }}{c^{2}}\frac{D_{LS}}{D_{OS}D_{OL}}\, ,
\end{equation}
 where \( D_{OL} \) is the observer-lens distance, \( D_{LS} \)
is the lens-source distance, and \( D_{OS} \) is the observer-source
distance%
\footnote{In flat spacetime, which is relevant for Galactic lensing, \( D_{OS}=D_{OL}+D_{LS} \).
In curved spacetime, which is relevant for cosmological lensing, the
distances are the angular diameter distances, and this simple sum
no longer holds. 
}.

The relation between the angular position of the source relative to
the observer-lens axis (the optical axis) can be derived from the
geometry of the light paths,

\begin{equation}
\label{eq:yx}
y=x_{1,2}-1/x_{1,2}\, ,
\end{equation}
 where \( x_{1,2} \) and \( y \) are measured in terms of \( \theta _{E} \)
and \( x_{2}<0 \) by definition. Gravitational lensing conserves
surface brightness, and so the magnifications \( A_{1,2} \) in the
flux of each image relative to that of the unlensed source is proportional
to change in the angular area of the source,

\begin{equation}
\label{eq:GLA}
A_{1,2}=\left| \frac{\partial \overrightarrow{y}}{\partial \overrightarrow{x}_{1,2}}\right| ^{-1}=\left| 1-x_{1,2}^{-4}\right| ^{-1}\, .
\end{equation}
 The primary image at \( x_{1} \) is always magnified. The secondary
image at \( x_{2} \) can be demagnified to zero. The two magnifications
obey the relations

\begin{equation}
\label{eq:A12}
A_{1}=A_{2}+1\geq 1\, ,
\end{equation}
 and

\begin{equation}
\label{eq:A}
A\equiv A_{1}+A_{2}=\frac{y^{2}+2}{y\sqrt{y^{2}+4}}\, .
\end{equation}
 When \( y=0 \) the amplification formally diverges and the image
appears as a ring of angular size \( \theta _{E} \), the Einstein
ring. This divergence is avoided in practice by the finite size of
the source (e.g. a star). Finite sized sources are also sheared tangentially
around the Einstein ring as the magnification increases. In the limit
of high magnification, or small source angle,

\begin{equation}
A\sim 1/y\quad (y\ll 1)\, .
\end{equation}

\begin{figure}[tp]
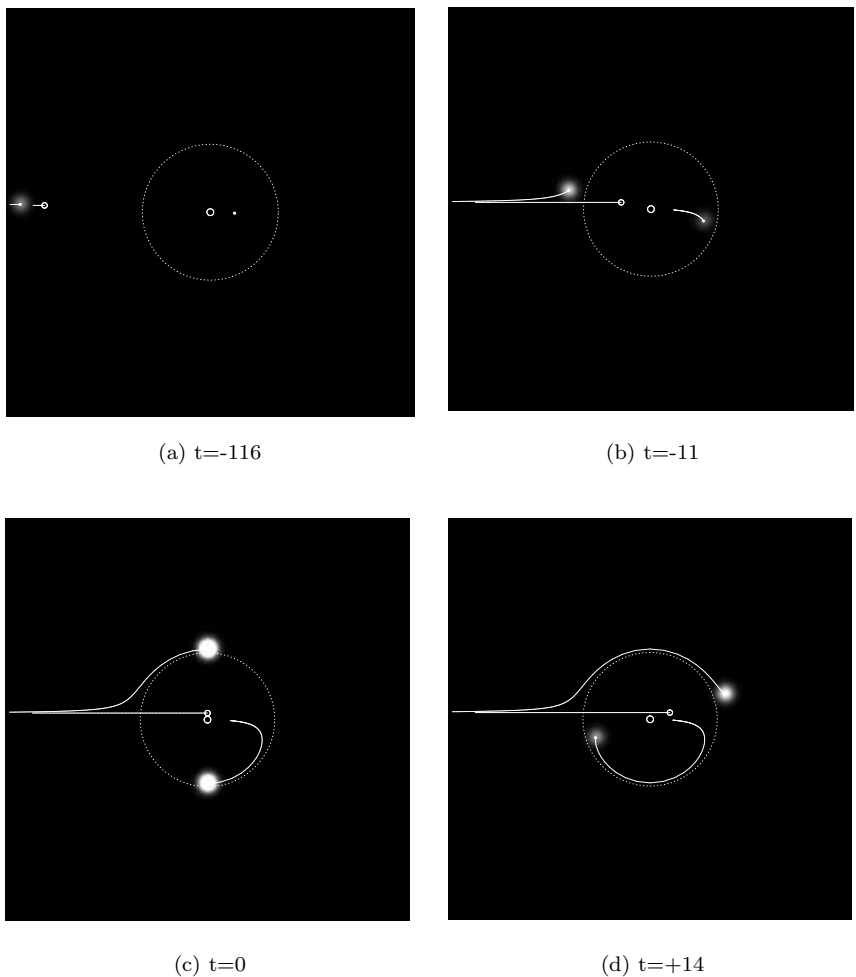

{\centering \vspace*{1mm}\par}

{\centering \begin{tabular}{cc}
\hline 
\subfigure[t=-116]{\resizebox*{0.48\textwidth}{!}{\rotatebox{270}{\includegraphics{gl2im_1.ps}}}} &
\subfigure[t=-11]{\resizebox*{0.48\textwidth}{!}{\rotatebox{270}{\includegraphics{gl2im_2.ps}}}} \\
\subfigure[t=0]{\resizebox*{0.48\textwidth}{!}{\rotatebox{270}{\includegraphics{gl2im_3.ps}}}} &
\subfigure[t=+14]{\resizebox*{0.48\textwidth}{!}{\rotatebox{270}{\includegraphics{gl2im_4.ps}}}} \\
\end{tabular}\par}

\caption{A sequence of snapshots simulating the observation of lensing of
a background point source by the MBH. Time is in arbitrary units.
The background source, which is not observed directly, (open circle
with straight line tracking the source trajectory) moves in projection
from left to right behind the MBH (open circle at center) with an
impact parameter of \protect\( 0.1\theta _{E}\protect \). The two
images (light points with curved lines tracking the image trajectories)
move in tandem clockwise about the Einstein ring (large dotted circle).
The strongly amplified image (top) is always outside the Einstein
ring and is always brighter than the source. The weakly amplified
image (bottom) is always inside the Einstein ring and can be strongly
deamplified (panels a, b). At peak amplification (panel c) the two
images are of comparable brightness (equation \ref{eq:A12}). }

\label{fig:GLimages}
\end{figure}

\subsection{Pinpointing the MBH with lensed images}

\label{sec:Pinpoint}

\begin{figure}[tp]
{\centering \vspace*{1mm}\resizebox*{1\textwidth}{!}{\rotatebox{270}{\includegraphics{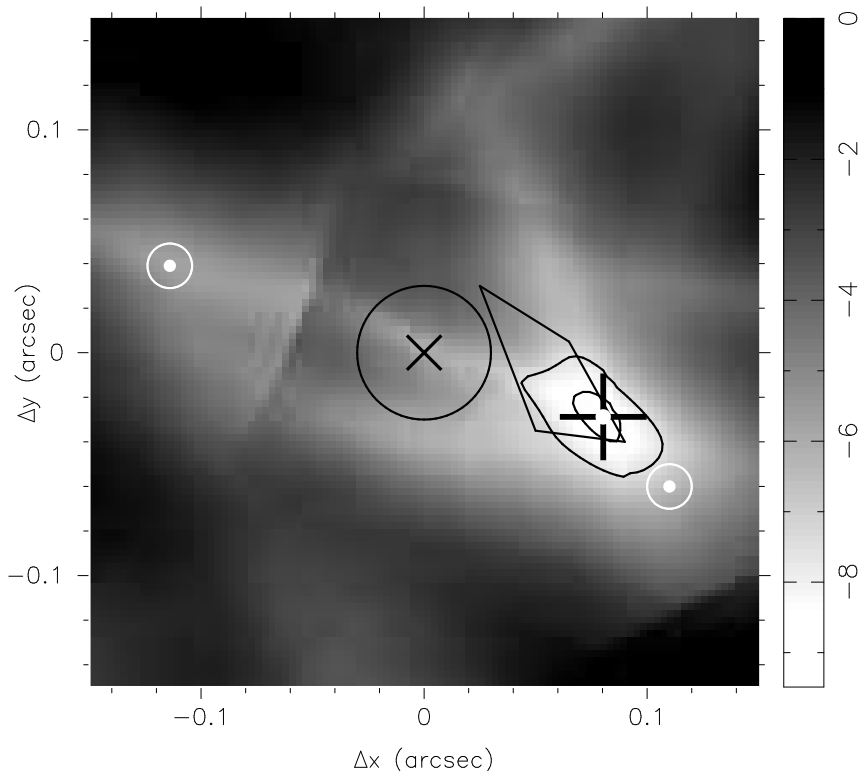}}} \par}

\caption{Pinpointing the MBH on the IR grid with gravitationally lensed stars
(Alexander \cite{Ale01a}). A gray scale plot of \protect\( \ln \mathrm{ML}\protect \)
(shifted to 0 at the maximum) for 116 stars from the astrometric compilation
by Genzel et al. (\cite{Gen00}), as function of the shift in the
astrometric grid over the central \protect\( 0.3''\times 0.3''\protect \)
search field. The cross in the center is the origin according to the
IR/radio alignment with its \protect\( 1\sigma \protect \) error
circle (Menten et al. \cite{Men97}). The polygon is the \protect\( \sim \! 1\sigma \protect \)
error region for the center of acceleration (Ghez et al. \cite{Ghe00}).
The circles are the observed IR sources with their 10 mas error circles.
The most likely position of the MBH is indicated by a plus sign with
1\protect\( \sigma \protect \) and 2\protect\( \sigma \protect \)
confidence level contours. (Reprinted with permission from \emph{The
Astrophysical Journal}).}

\label{fig:Pinpoint}
\end{figure}

Determining the exact position of the MBH on the IR grid is important
because the radio source SgrA\( ^{\star } \), which is associated
with the MBH, was detected to date only in one other band, the X-ray
(Baganoff et al. \cite{Bag01}). Currently, the IR position of the
radio source SgrA\( ^{\star } \) is derived indirectly by aligning
the radio and IR maps using 4 maser giants in the inner \( 15'' \),
which are observed in both bands (Menten et al. \cite{Men97}). The
exact IR position of the MBH is required, for example, for measuring
the IR flux from the MBH, in order to constrain accretion models;
for solving the stellar orbits around the MBH, in order to measure
\( M_{\bullet } \) and \( R_{0} \) (Jaroszy\'{n}ski \cite{Jar99};
Salim \& Gould \cite{Sal99}) and search for general relativistic
effects (Jaroszy\'{n}ski \cite{Jar98}; Fragile \& Mathews \cite{Fra00};
Rubilar \& Eckart \cite{Rub01}); and for detecting the fluctuations
of the MBH away from the dynamical center of the stellar cluster,
in order to study the stellar potential. Recent measurements of the
acceleration vectors of three stars very near SgrA\( ^{\star } \)
provide another way of locating the MBH (Ghez et al., \cite{Ghe00}).
The IR/radio alignment and the center of acceleration are close, but
do not overlap (Figure \ref{fig:Pinpoint}), and neither coincide
with an IR source. Gravitational lensing can provide a third, independent
method for locating the MBH.

When the source, lens and observer move relative to each other, the
positions, velocities and magnifications of the images will change
with time (Figure \ref{fig:GLimages}). In addition to the requirement
that the two images and the lens lie on one line, equations (\ref{eq:yx})
and (\ref{eq:GLA}) imply that the measured angular positions of the
two images \( \theta _{1,2} \), their projected transverse velocities
\( v_{t1,2} \) and radial velocities \( v_{r1,2} \) relative to
the lens, and their measured fluxes \( F_{1,2} \), should obey the
simple relation

\begin{equation}
\label{eq:pinpoint}
-\theta _{1}\left/ \theta _{2}\right. =v_{t1}\left/ v_{t2}\right. =-v_{r1}\left/ v_{r2}\right. =\sqrt{F_{1}\left/ F_{2}\right. }\, .
\end{equation}
 The constraints are based solely on observables, and so are independent
of any assumptions about \( M_{\bullet } \), \( R_{0} \) or the
properties of the lensed background sources. The use of equation (\ref{eq:pinpoint})
does require knowledge of the exact position of the MBH relative to
the stars, since this is needed for measuring the angular distances
and for decomposing the radial and tangential components of the velocity.
If the MBH position is known, equation (\ref{eq:pinpoint}) can be
used to search in astrometric measurements of positions, fluxes and
velocities for pairs of lensed images around the MBH. Equation (\ref{eq:pinpoint})
can also be used to find the position of the MBH on the IR grid, since
the MBH lies on the line connecting the two images, and so the intersection
of these lines pinpoints its position. This can be done statistically,
by enumerating over a grid of trial positions for the MBH, and choosing
as the most likely one that which maximizes the number of lensed image
pairs.

Figure (\ref{fig:Pinpoint}) shows the results from such a joint statistical
search for the MBH and for a signature of lensing (Alexander \cite{Ale01a}).
The most likely position of the MBH coincides with the center of acceleration.
The random probability for such a likelihood extremum is \( 0.01 \).
The random probability for such an extremum to fall in either the
\( 1\sigma  \) error range of IR/radio alignment or that of the center
of acceleration is \( 5\times 10^{-4} \).

The search for the MBH yields also a list of candidate lensed image
pairs. The definitive test of lensing is to compare their spectra,
which should be identical up to differences due to non-uniform extinction.
Unfortunately, spectra for the fainter secondary images are unavailable
at this time. Once \( M_{\bullet } \), \( R_{0} \) and the dust
distribution in the Galaxy are assumed, it is possible to derive,
albeit with very large uncertainty, the luminosity and distance of
the candidate sources. The sources of the two most likely lensed image
pair candidates are luminous supergiants, a blue supergiant a few
kpc behind the Galactic Center and a red supergiant at the far edge
of the Galaxy.

This statistical result, while intriguing, requires additional confirmation.
Simple models of the distribution of light and dust in the Galaxy
predict that the chances of finding luminous supergiants right behind
the MBH are very small, and the statistical analysis depends sensitively
on the quality of the data and its error properties. Whether or not
this particular result survives further scrutiny, it illustrates the
potential of gravitational lensing as a tool for the study of the
Galactic Center. This statistical method for locating the MBH by gravitational
lensing should be re-applied whenever deeper astrometric data become
available.

\subsection{The detection of gravitational lensing}

The mode of detection of lensing events depends on the telescope's
spatial resolution and its photometric sensitivity. When the two images
can be resolved, as in the case discussed in \S\ref{sec:Pinpoint},
the phenomenon is called a {}``macrolensing'' event. When the two
images cannot be resolved, only the variability in the flux of the
lensed source is observed. This is called a {}``microlensing'' event.
Since \( \theta _{E} \) increases with source distance behind the
lens, there is a maximal source distance for microlensing, \( D_{\mu } \),
which can be estimated by noting that the angular distance between
the two images close to peak magnification is \( \sim \! 2\theta _{E} \),
and so

\begin{equation}
D_{\mu }=\frac{D_{OL}}{\left( \theta _{\infty }/\phi \right) ^{2}-1}\, ,
\end{equation}
 where \( \phi  \) is the telescope's angular resolution, \( \theta _{\infty }\equiv \sqrt{4GM_{\bullet }\left/ c^{2}D_{OL}\right. }\sim 1.75^{''} \)
is the Einstein angle for a source at infinity, and it is assumed
that \( \phi <\theta _{E} \) is the criterion for resolving the two
images.

\begin{figure}[tp]
{\centering \vspace*{1mm}\resizebox*{1\textwidth}{!}{\rotatebox{270}{\includegraphics{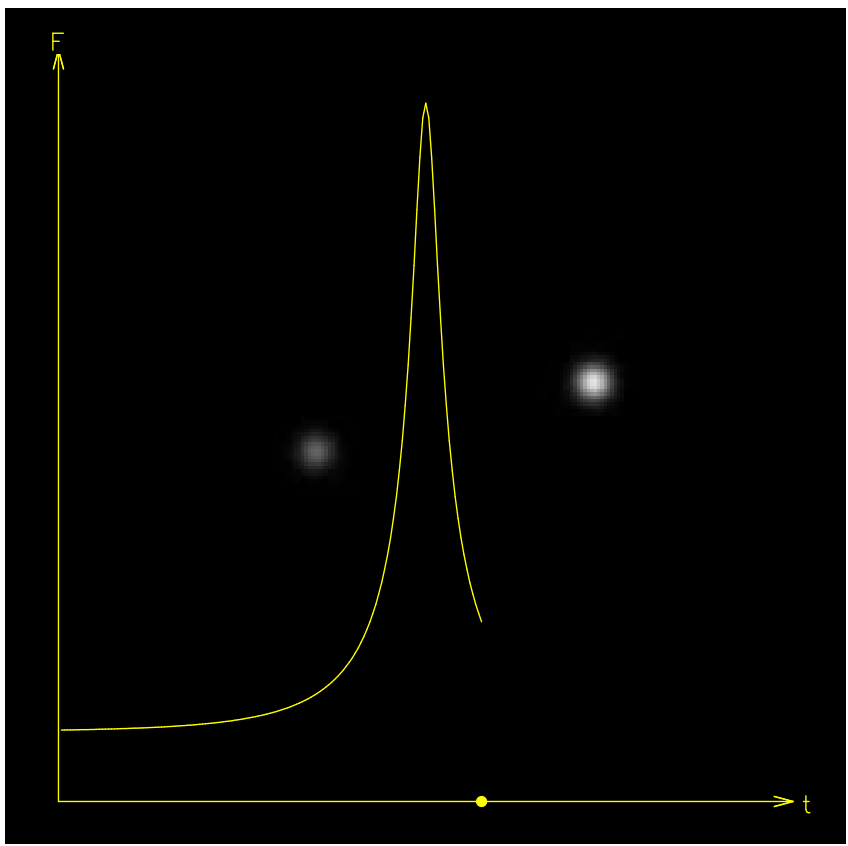}}} \par}

\caption{The microlensing light curve that corresponds to the lensing event
in Figure \ref{fig:GLimages} up to time \protect\( t=19\protect \),
in the case where the two images can not be resolved. The light curve
is overlayed on the two lensed images as they would be observed if
they could be resolved. The point on the time axis indicates the position
of the source along its trajectory. The flux level on the left is
the unmagnified flux of the source. The complete light curve will
be symmetric relative to the peak flux.}

\label{fig:GLlc}
\end{figure}

The light curve for a constant velocity trajectory of a background
source in the plane of the sky is given by substituting \( y(t) \)
in equation (\ref{eq:A}), \begin{equation}
y^{2}(t)=y^{2}_{0}+\mu ^{2}(t-t_{0})^{2}\, ,
\end{equation}
 where \( y_{0} \) is the impact parameter of the source trajectory
relative to the lens, \( \mu  \) is the apparent motion, in units
of \( \theta _{E} \) per time, and \( t_{0} \) is the time when
\( y=y_{0} \) (Figure \ref{fig:GLlc}). The resulting light curve
is symmetric about \( t_{0} \), and achromatic (i.e. has the same
shape in every wavelength). If the photometric sensitivity is large
enough to detect the unlensed source, the event will appear as a flaring
up of a persistent source, otherwise, it will appear as a transient
flare.

In order to plan the observational strategy for detecting gravitational
lensing, or to estimate how likely it is that an observed flare is
due to lensing, it is necessary to calculate the detection probability.
Two quantities are commonly used to express this probability, the
optical depth and the lensing rate. The optical depth for gravitational
lensing is usually defined in relation to the probability of having
at least one lens along the line of sight. If the cross-section of
the lens at position \( z_{i} \) is \( S(z_{i}) \), and the number
density of lenses there is \( n(z_{i}), \) then the probability \( P \)
of having at least one lens along the line of sight is the complement
of the probability of not encountering any lens,

\begin{equation}
\label{eq:GLtau}
\begin{array}{ccccc}
P & = & 1-\prod _{i}(1-n(z_{i})S(z_{i})\Delta z_{i}) &  & \\
 & = & 1-\exp (-\int ^{z}_{0}nSdz^{\prime }) &  & \\
 & = & 1-e^{-\tau } & \rightarrow  & \tau \qquad (\tau \ll 1)
\end{array}
\end{equation}
 It should be emphasized that \( \tau  \) is \emph{not} a probability,
and that \( P \) and \( \tau  \) are interchangeable only when both
are small. It is customary to define the lensing cross-section as
\( S=\pi \theta _{E}^{2} \), that is, the region where a source will
be amplified by \( A>1.34 \) (equation \ref{eq:A12}). This definition
is useful when there are many possible lines of sights, and it describes
the probability that at \emph{any given instant} a given line of sight
will be lensed. This is relevant for Galactic microlensing searches,
where millions of background stars are monitored simultaneously to
find the rare one that is lensed by an intervening star. The observational
situation for gravitational lensing by the MBH in the Galactic Center
is different because the position of the lens is known and fixed,
and so there is only one line of sight. In analogy to equation (\ref{eq:GLtau}),
the optical depth is defined in relation to the probability of having
at least one source behind the lens along the line of sight,

\begin{equation}
\begin{array}{ccc}
\tau  & = & \int ^{\infty }_{D_{OL}}n_{\star }\pi R_{E}^{2}dD_{OS}
\end{array}\, ,
\end{equation}
 where \( R_{E} \) is the physical size of \( \theta _{E} \) at
the source plane,

\begin{equation}
\label{eq:RE}
R_{E}=\theta _{\infty }\sqrt{\left( D_{OS}-D_{OL}\right) D_{OS}}\, .
\end{equation}
 Rough estimates predict \( \tau \! \sim \! 1 \) for lensing by the
MBH (assuming no limits on the photometric sensitivity).

The optical depth does not take into account the relative motions
of the lens and source, which reshuffle their random alignment and
introduces a timescale to the problem. A more useful quantity for
the lensing by the MBH is the lensing event rate with flux above a
detection threshold \( F_{0} \) due to the motion of sources behind
the MBH,

\begin{equation}
\label{eq:GLrate}
\Gamma (>F_{0})\simeq 2\int _{D_{OL}}^{\infty }n_{\star }v\frac{R_{E}}{A}dD_{OS}\, ,\qquad A\geq \frac{F_{0}}{\left. L_{\star }\right/ 4\pi D^{2}_{OS}}\, ,
\end{equation}
 where \( v_{\star } \) is the source star's projected velocity,
\( L_{\star } \) is its luminosity and \( A\gg 1 \) is assumed.
For practical applications, equation (\ref{eq:GLrate}) has to be
modified to take into account the range of stellar luminosities and
velocities, dust extinction, the total duration of the observations
\( T \) and the sampling rate \( \Delta T \) (the mean duration
of events amplified by more than \( A \) is \( \overline{t}=\pi R_{E}\left/ 2Av\right.  \);
only events with \( \Delta T<t<T \) can be detected).

\begin{figure}[tp]
{\centering \vspace*{1mm}\resizebox*{0.9\textwidth}{!}{\includegraphics{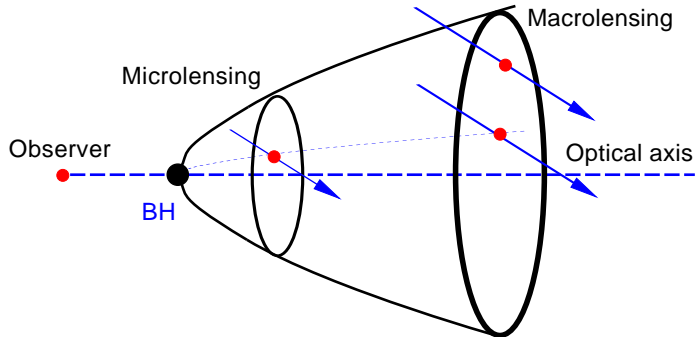}} \par}

\caption{The lensing cross section, timescale and amplification as function
of source distance behind the MBH. The size of the Einstein ring,
or lensing cross-section, (cone, see equation \ref{eq:RE}) increases
with distance behind the MBH. Close behind the MBH The Einstein ring
(and the distance between the two images) is smaller than the telescope's
resolution and the lensing appears as a microlensing event. Farther
out, the Einstein ring is large enough for the two images to be resolved,
and the lensing appears as a macrolensing event. The duration and
peak magnification of the events depend on the impact parameter of
the stellar trajectories (arrows). The closer they are to the optical
axis, the longer the events and the higher the peak magnification.
Trajectories with impact parameters at a fixed ratio of the Einstein
radius (the two trajectories connected by the dotted line) will have
the same peak amplification (equation \ref{eq:A}), but the event
duration will be longer for the sources farther away behind the MBH
(assuming a uniform velocity field).}

\label{fig:GLxsection}
\end{figure}

Figure \ref{fig:GLxsection} summarizes the dependence of the lensing
cross-section, timescale and amplification on \( D_{LS} \). The observational
limitations, \( F_{0} \), \( T \) and \( \Delta T \), place restrictions
on \( D_{LS} \) and the impact parameter for which sources can be
detected, and affect the typical timescales and peak magnification
that are likely to be observed. For example, high magnification events
typically have longer time scales because the source trajectory must
have a smaller impact parameter and so spends more time in the Einstein
radius. Therefore, observations with limited temporal sampling will
tend to pick out high magnification events.

Were any microlensing events detected? A couple of possible transient
flaring events were detected very close to SgrA\( ^{\star } \) (Genzel
et al. \cite{Gen97}; Ghez et al. \cite{Ghe98}). For one of these
a light curve was recorded, but as it was under-sampled only estimates
of a timescale (\( \sim \! 1 \) yr) and a typical magnification (\( \sim \! A>5 \))
could be derived from it. The a-posteriori probability of detecting
a lensing event was estimated at only \( 0.5\% \), but on the other
hand, the observed timescale and magnification are close to the median
value that is expected for the observational limitations (Alexander
\& Sternberg \cite{Ale99b}). The interpretation of this event remains
inconclusive.

\subsection{Magnification bias}

\begin{figure}[tp]
{\centering \vspace*{1mm}\resizebox*{0.9\textwidth}{!}{\includegraphics{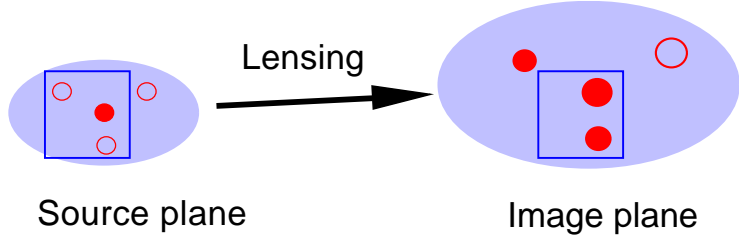}} \par}

\caption{The magnification bias in star counts due to gravitational lensing.
Stars (circles) are counted in a field of a fixed angular area (square)
with a telescope of a given photometric detection threshold. The field
in the unlensed sky (source plane, left), contains one bright star
that can be observed (filled circle) and two stars that are too faint
to be observed (open circles). Gravitational lensing stretches angular
areas and amplifies fluxes by the same factor. The field in the lensed
sky (image plane, right), now contains only two stars, but because
they are magnified, both can be observed. This is an example of positive
magnification bias, where gravitational lensing increases the \emph{apparent} stellar
surface density even as it decreases the \emph{total} surface density.
Negative magnification bias occurs when there aren't enough faint
stars in the lensed population to compensate for the decrease in the
total surface density. }

\label{fig:GLbias}
\end{figure}

A lens magnifies by enlarging the angular size of the unlensed sky
behind the lens, and since surface brightness is conserved, the fluxes
of sources are magnified by the same amount. When the photometric
sensitivity is such that all the stars can be detected even without
being magnified, then the effect of lensing is to decrease the surface
density of sources. However, if the fainter stars cannot be observed
unless magnified, there are two possibilities (Figure \ref{fig:GLbias}):
either there are enough faint sources that are magnified above the
detection threshold to over-compensate for the decrease in surface
density ({}``positive magnification bias''), or there aren't enough
faint sources ({}``negative magnification bias''). The lensed luminosity
function (number of stars per flux interval) is related to the unlensed
one by

\begin{equation}
\label{eq:GLbias}
\left. \left( \frac{d\Sigma }{dF}\right) _{\mathrm{lensed}}\right| _{F}=A^{-2}\left. \left( \frac{d\Sigma }{dF}\right) _{\mathrm{unlensed}}\right| _{F/A}\, ,
\end{equation}
 where \( \Sigma  \) is the surface number density of stars and \( F \)
the flux. In many cases the luminosity function is well approximated
by a power-law, \( d\Sigma \left/ dF\right. \propto F^{-\beta } \).
It then follows from equation (\ref{eq:GLbias}) that for \( \beta =2 \)
the decrease in the total surface density is exactly balanced by the
magnification of faint stars above the detection threshold.

The chances for the detection of this effect in the Galactic Center
appear small. A statistically meaningful detection requires a very
high surface density that probably exceeds even that around the MBH
(Wardle \& Yusef-Zadeh \cite{War92}), and furthermore, models of
the stellar luminosity function in the inner Galactic Center suggests
that \( \beta \sim 2 \) for giants (Alexander \& Sternberg \cite{Ale99b}).

\subsection{Beyond the point mass lens approximation}

Up to this point we considered only the simple case of lensing by
a point mass. There are two reasons to explore more complicated models.
The first is that it would be useful if gravitational lensing could
be used to dispel any remaining doubts that the dark compact mass
in the Galactic Center is indeed a MBH, and not some other extended
distribution of matter, such as an compact cluster of stellar remnants
(Maoz \cite{Mao98}) or a concentration of exotic particles (Tsiklauri
\& Viollier \cite{Tsi98}). Unfortunately, it can be shown the behavior
of high-magnification light curves near peak magnification is universal
and independent of the details of the lens (equation 11.21b of Schneider,
Ehlers \& Falco \cite{Sch92}). For spherically symmetric mass distributions
this implies that the light curves differ only in the low magnification
tails, which are much harder to observe. The second reason is that
the MBH is surrounded by a massive stellar cluster. Because the stellar
mass is not smoothly distributed but is composed of discrete point
masses, its effect on the lensing properties of the MBH is much larger
than one may naively estimate by adding the stellar mass to that of
the MBH. We conclude the discussion of gravitational lensing in the
Galactic Center with describing briefly the effect of enhanced lensing
by stars near the MBH (Alexander \& Loeb \cite{Ale01d}; Chanam\'{e},
Gould \& Miralda-Escud\'{e} \cite{Cha01}).

\begin{figure}[tp]
{\centering \vspace*{1mm}\resizebox*{0.9\textwidth}{!}{\includegraphics{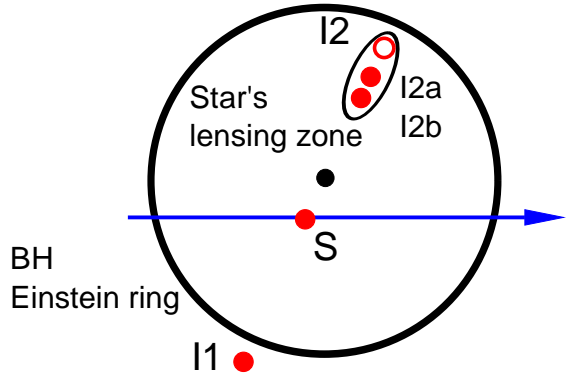}} \par}

\caption{A schematic representation of lensing enhancement by a star near
the MBH (Alexander \& Loeb \cite{Ale01d}; Chanam\'e, Gould \& Miralda-Escud\'e
\cite{Cha01}). A light source S passes behind the MBH (filled circle
in the middle) and, in the absence of any other lensing mass, appears
as two images: I1 outside the Einstein ring and I2 inside the Einstein
ring. When one of the stars near the MBH happens to lie (in projection)
close to I2, it will split I2 into two or four sub-images (here shown
two), I2a and I2b. The star's Einstein ring is sheared by the potential
of the MBH to an elongated shape of complex topology (represented
here for simplicity as an ellipse), which increases in size the nearer
I2 and the star are to the Einstein radius of the MBH. This effect
increases the cross-section for high magnification events above that
of an isolated MBH, and changes the character of the light curves. }

\label{fig:GLbinary}
\end{figure}

The effect of stars on lensing by the MBH is similar to that of planets
on microlensing by a star, an issue that was studied extensively for
the purpose of detecting planets by microlensing (e.g. Gould \& Loeb
\cite{Gou92}). The lensing cross-section of an isolated star is \( \theta ^{2}_{E}(M_{\star })/\theta ^{2}_{E}(M_{\bullet })=M_{\star }/M_{\bullet }\lesssim 10^{-6} \)
smaller than that of the MBH (Equation \ref{eq:ThetaE}). However,
when the star lies near \( \theta (M_{\bullet }) \), the shear of
the MBH distorts its cross-section, which develops a complex topology,
becomes radially elongated and is increased by up to an order of magnitude
(Figure \ref{fig:GLbinary}). As the stars orbit the MBH, their elongated
cross-sections scan the lens plane. If these happen to intersect one
of the images of a background source that is lensed by the MBH, the
image will be split into 2 or 4 sub-images whose angular separation
will be of order \( \theta _{E}(M_{\star }) \), and so the sub-images
will not be individually resolved. However, their combined flux will
be significantly magnified. This will increase the probability of
high magnification events over what is expected for lensing by the
MBH alone. The light curves of such events will no longer be symmetric
as they are for a point mass, but will exhibit a complex structure
(e.g. Wambsganss \cite{Wam97}), and their typical variability timescales
will rise sharply for images that lie near \( \theta _{E}(M_{\bullet }) \)
because of the increased stellar cross-section for lensing. Enhanced
lensing by stars in the Galactic Center is estimated to increase the
probability of \( A>5 \) lensing events by \( \sim \! 2 \) and of
\( A>50 \) events by \( \sim \! 3 \).

\section{Summary}

\label{sec:Summary}

Observations of the MBH in the Galactic Center present a unique opportunity
to study the consequences of extreme stellar density, velocity and
tidal fields on the dynamics and evolution of stars and their relation
to the dynamics and evolution of the MBH. The existence of a high
density relaxed stellar cusp around the MBH in the Galactic Center
is theoretically motivated, and supported by observations. We explored
some of the consequences of this environment for the appearance, internal
structure and evolution of stars, through exotic object formation
by direct collisions, collisional destruction of giant envelopes,
stochastic tidal spin-up of stars by collisions with other stars,
and extreme tidal interactions in the course tidal scattering by the
MBH. It was shown that tidal processes have the potential of affecting
a significant fraction of the stars over a large volume around the
MBH.

The MBH is also a gravitational lens. This can be used to probe the
dark mass and the stars around it, but it also has the potential for
complicating the interpretation of observations in the Galactic Center.
Different detection modes were considered: macrolensing, microlensing,
magnification bias, and the detection probability and detection rate
were defined. Results from a statistical method for detecting lensed
images and for pinpointing the MBH on the IR grid suggest that there
may be a few far background supergiants that are lensed by the MBH.
We described a lensing effect that involves both the MBH and the stars
around it, and can increase the probability of high magnification
events and modify the structure of the light curves.

The topics covered by this chapter by no means exhaust the scope of
the subject. We did not address, among others, star formation near
the MBH, the role of stellar evolution in feeding the MBH, or compact
stellar remnants and x-ray sources. Some of these issues are discussed
elsewhere in this book.

Over the next decade a wide array of IR instruments, both ground based
and space borne, will improve the quality of photometric, spectroscopic
and astrometric observations of the Galactic Center by orders of magnitude.
Many of the issues discussed here will be resolved, as new questions
will surely be raised. One thing is certain---we can look forward
to exciting times in Galactic Center research.

\end{document}